\documentclass[final,3p,times,twocolumn]{elsarticle}
\usepackage{graphicx}
\usepackage{amssymb}
\usepackage{amsmath}
\usepackage{hyperref}
\usepackage{longtable}
\usepackage{subfig}

\biboptions{sort&compress}
\journal{Annals of Physics}

\newcommand{\preprint}{
\setlength{\unitlength}{1mm}{\hbox{\begin{picture}(0,0)
      \put(160,10){\mbox{\footnotesize%
        ADP-10-22/T718}}\end{picture}}}}
\begin{document}
\begin{frontmatter}

\title{\preprint On the ground state of Yang-Mills theory}
\author{Ahmed S. Bakry}
\ead{abakry@physics.adelaide.edu.au}
\author{Derek B. Leinweber}
\author{Anthony G. Williams}

\address{Special Research Center for the Subatomic Structure of Matter, Department of Physics,\\ University of Adelaide, South Australia 5005, Australia}
\begin{abstract}  
  We investigate the overlap of the ground state meson potential with sets of mesonic-trial wave functions corresponding to different gluonic distributions. We probe the transverse structure of the flux tube through the creation  of non-uniform  smearing profiles for the string of glue connecting two color sources in Wilson loop operator. The non-uniformly UV-regulated flux-tube operators are found to optimize the overlap with the ground state and display interesting features in the ground state overlap. 
\end{abstract}


\end{frontmatter}

\section{Introduction}      

   A fundamental property of the non-perturbative regime of confining pure-gauge theories is the linear increase in the ground state potential between a pair of static color sources. In addition to that, lattice gauge simulations have recently confirmed the existence of a sub-leading non-Coulombic long range correction to the mesonic ground-state potential in Yang-Mills theory \cite{luscher,Juge:2002br,Caselle:1995fh,2004JHEP10005C}. These features are connected with the underlying gluonic picture and the subsequent energy distribution profile. Although the properties of the ground state potential have been unambiguously measured to a subleading order in the infrared region of the non-abelian gauge, the geometrical aspects of the associated energy density profile at low temperatures remains to be completely resolved. 

    Lattice calculations of the gluon-field distribution in static mesons using Wilson loop operators reveal uniform energy and action density profiles along the line joining the static $q\overline{q}$ pair at large distances ~\cite{Bali}. 
    These measurements, however, may be vulnerable to systematic errors associated with excited-state contamination ~\cite{Okiharu2004745}. The non-ground state components manifest themselves in the revealed gluonic profiles as a bias reflecting the form of spatial links of the Wilson loop operator. The bias by the geometry of spatial links in the $L$ shape baryon operator provides a clear example where the flux distribution mimics the source ~\cite{Okiharu2004745,Bissey}. The height and the width of the distribution also depend on the ultraviolet properties of the gauge links in the source \cite{Bissey}. Apart from the arbitrariness in adopting the source that best approximates the ground state, the statistical fluctuations impose a practical constraint on the Euclidean time evolution in the loop operator to isolate the physically interesting energy-density profile of the ground state. The excited-state contamination is more challenging in the case of field distribution calculations which involve three-point correlations rather than the potential which is extracted in the large time limit of a two point correlation \cite{Okiharu2004745}.  

\noindent In the finite-temperature regime, the static meson can be constructed using a pair of Polyakov lines. These  hadronic operators provide a systematically unbiased stringless gauge-invariant objects in the calculations of field-distribution correlations. This means that one need not adopt specific geometric or UV properties for the gluonic string between the color sources. 

  Current investigations of the flux-tube profile in the finite-temperature regime of QCD have revealed action-densities of non-uniform distribution along the flux-tube \cite{PhysRevD.82.094503}. The action density displays a two dimensional Gaussian-like profile and isosurfaces of a curved prolate spheroid-like shape \cite{PhysRevD.82.094503} in the intermediate source separation distance region $ 0.6 \leq R \leq 1 $ fm. This has been observed near the deconfinement point $ T \approx 0.9 T_c $ and remain manifest at the temperature $ T \approx 0.8 T_c $ close to the end of the plateau region of the QCD phase diagram \cite{Doi2005559}. The measurements of the tube's mean square width profile indicate, however, almost constant width topology. Variation in the amplitude give rise to curved isosurfaces. At larger distances, the tube changes width along the $q\overline{q}$ plane and this width profile is predictable based on a free bosonic string picture \cite{PhysRevD.82.094503}. The gluonic distributions obtained at finite temperature by correlating two Polyakov lines constitute an interesting source of knowledge for investigating the possibility that non-uniform densities provide the true geometry of the ground state in the static meson. The viability of considering finite temperature results as an indication for the field distribution of the system's ground state can be justified by arguing also that the change in the string tension is small \cite{kac,Bakry:2010sp} at $T\approx 0.8 T_{c}$.
 
  In addition to this observation, it has been found recently that a model of Coulombic trial states provides a good overlap  with the ground state in the continuum limit \cite{heinzl-2008-78}. Moreover, the free bosonic string model predicts  observable edge-effects at zero temperature for the width profile of the tube given by \cite{Allais} 

\begin{equation} 
  \dfrac{1}{\pi\sigma} \log \bigl|\cos(\dfrac{\pi \xi}{R})\bigl|, \xi \in [-R/2,R/2]
\end{equation}   

\noindent The above term describes the geometrical shape of the flux tube and it indicates subtle changes in the tube's mean square radius in the middle of the tube and  more pronounced changes near the quark positions. The success of the string picture in accounting for the flux-tube curvature at high temperature at large distances is remarkable, and one may investigate such effects at zero temperature. Apart from the string's width effects, a non-uniform action density amplitude pattern along the tube has been observed at finite temperature whether the tube exhibits a non constant width profile or not \cite{PhysRevD.82.094503}. The Bag model is also anther scheme that predicts an ellipsoidal-like \cite{Juge1998543} shape for the tube in the infrared region.  
             

 At zero temperature, correlating a pair of Polyakov lines with an action density operator is very noisy and requires substantial numerical simulations using special techniques such as the Multi-level algorithm \cite{luscher}. Nevertheless, standard Wilson loop operators do not exhaust the possibility of investigating the transverse structure of the field distribution. We can introduce the non-uniformity by employing the idea of constructing the flux-tube operator as a product of locally smeared links with varying smearing extents. This corresponds to imposing a local transverse cutoff on the lattice parallel transporters between the fermionic fields. By extending the space of mesonic states constructed in Wilson loop this way, we investigate the existence of states such that the overlap with the ground state is maximized. This is the objective of the present report.  

\section{Wilson loop operator}
In the mesonic Wilson loop, a mesonic state is described by fermionic fields connected by the parallel transporters $\mathcal{G}$ that is an element of the corresponding local gauge group $G$ 
\begin{equation}
 \vert \Psi \rangle= \bar{\psi}(x_2) \, \mathcal{G} \psi(x_1)\, \vert \, \Omega \, \rangle .
\end{equation}

\noindent The spectral expansion of the Wilson loop operator reads
\begin{equation}
  \langle W(R,t) \rangle= \sum_n C_n(R) e^{-V_n(R)t} .
\end{equation}  

\noindent The overlaps, $C_n(R)$, obey the normalization condition

\begin{equation}
  \sum_n C_n(R)=1.
\end{equation}

\noindent For large $t$, the so-called overlap with the system ground state can be measured as

\begin{equation}
  C_0(R) = \dfrac{ \langle W(R,t) \rangle^{t+1}} {\langle  W(R,t+1) \rangle^{t}}. 
\label{over}
\end{equation}
The mesonic state with an infinitesimally thin flux tube operator between quarks
\begin{equation}
\mathcal{G}=\mathcal{P} \exp \Big[ \int_{x_{2}}^{x_{1}}\, d\bf{z\,.\,A(z)} \Big ]
\end{equation}
\noindent is  poorly overlapping with the ground state in the continuum \cite{heinzl-2008-78}. Analogy with an Abelian analytically solvable case \cite{Heinzl:2007kx} shows that the infinitesimal thickness of the gauge links corresponds to the removal of UV cutoff on the transverse direction of the tube causing a vanishing overlap with the ground state. 

\begin{figure}[!hpt]
\begin{center}
\includegraphics[ width=8cm]{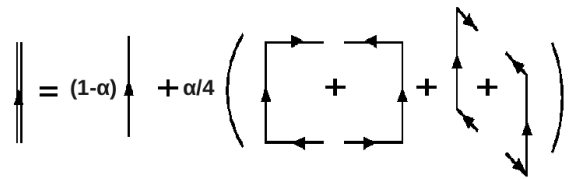}
\caption{\label{LB}Schematic representation of link-blocking}
\end{center}
\end{figure}

\noindent  Applying a smearing operator on a gauge link in a certain local gauge group would alter its UV properties. Smearing introduces a transverse UV regulator into the flux-tube operator and this results in the enhancement of the overlap with the ground state. 
Define a smearing operator $\mathcal{S}:\oplus_xG \to \oplus_xG$. A standard APE \cite{Albanese}smearing sweep ($\mathcal{S}U)_\mu(x)$ consists of a replacement of the spatial link-variable $U_\mu (x)$ ($\mu=1,2,3$) with the angular part of 
\begin{align}
 U_{s,\mu}(x)=&(1-\alpha) \,U_\mu(x)\notag +\frac{\alpha}{4} \sum_{\mu \ne \nu } \{U_\nu(x)U_\mu(x+\hat \nu) \notag \\
                       & U_\nu^\dagger (x+\hat \mu) + U_\nu^\dagger (x-\hat \nu)U_\mu(x-\hat \nu)U_\mu(x+\hat \nu-\hat \mu)\}, 
\label{APE}
\end{align}
\noindent where $\alpha$ is the smearing parameter as illustrated Fig.~\ref{LB}. In QCD, this corresponds to the projected link $\bar U_\mu (x)  \in{\rm SU(3)}_c$ that maximizes  
\begin{equation}
\Re \, \mathrm{Tr} \{ \bar U_\mu(x) U_{s,\mu}^{\dagger}(x) \} .
\label{proj}
\end{equation}
\noindent The geometrical characteristics of smearing can be described by  analogy to the Brownian motion associated with diffusing a scalar field \cite{Takahashi}. Given a scalar field $\phi({\bf r};n+1)$ similar to the \mbox{$(n+1)$-th} smeared gauge link in the $\mu$ direction, and  a smearing time $ \tau= n \, a_{\tau}$ with a spacing $a_{\tau}$. The smearing operation will then correspond to the diffusion initial value problem,

\begin{align}
\partial_{\tau}  \phi({\bf r};\tau)= D \, (\partial^{2}_x+&\partial^{2}_y)\, \phi({\bf r};\tau),\notag \\
\phi({\bf r};n=0)= & \delta({\bf r}) \notag \\
\intertext{with the diffuseness,} 
D \equiv \frac{\alpha}{4} \frac{a^2}{a_\tau}.
\label{Diff}
\end{align}%
\noindent The Green kernel of the above heat equation partial differential equation, Eq.~\eqref{Diff}, gives the evolution of the scalar field in the smearing time, 
\begin{equation}
	G({\bf r};\tau)=
	{1\over (4\pi D  \tau)}
	\exp\left[ - {{\bf r.r} \over 4 D  \tau} \right].
\label{green}
\end{equation}
\noindent The diffuse field is Gaussian distributed with a characteristic radius,
\begin{align}
r \equiv& \left(\frac{\int d^{3}{\bf r} \, G({\bf r};\tau)   {\bf r}^{2} }
{\int d^{3}{\bf r} \, G({\bf r};\tau)}\right)^{1/2} =2\, a \sqrt{\alpha \, n }.
\label{size}
\end{align}

\noindent Applying the local smearing operator $\mathcal{S}$ at each spatial link in Wilson loop operator,
\begin{equation}
  \mathcal{G} =  (\mathcal{S}^{n_1}U)(x_1)\, (\mathcal{S}^{n_2}U)(x_1+a) \cdots 
\label{gdef}
\end{equation}
\noindent The sequence of numbers of smearing sweeps applied at each link $\{n_1,n_2,n_3,...,n_N\}$ fixes the gluonic distribution along the spatial links in Wilson loop. This sequence of numbers maps into the geometrical space of the corresponding radii $r(x_{i})$ and the amplitudes $A(x_{i})$ given by Eqs.~\eqref{green} and \eqref{size}. For a mesonic state ${n_{i}}$ or $r(x_{i})$, the  projection on the system's ground state is measured as, 
\begin{equation}
  \langle  \Psi_{0} \vert \Psi_{\{n\}} \rangle =\langle  \Psi_{0} \vert \bar{\psi}(\bf{x_2}) \mathcal{G} \psi(x_1) \vert \Omega \rangle . 
\end{equation}


\noindent The mesonic state constructed by operators corresponding to a rectangular shape given by a constant sequence $\{r_i\}$ has been considered to provide a good approximation for the potential ground state \cite{Bali}. The understanding of the geometry of the flux-tube, nevertheless, would be increased by constructing trial states without this constraint. We expect based on the results in Refs.~\cite{heinzl-2008-78,PhysRevD.82.094503} that the best possible approximation of the ground state may be approached this way.

\begin{figure}[!hpt]
\begin{center}
\includegraphics[width=8cm]{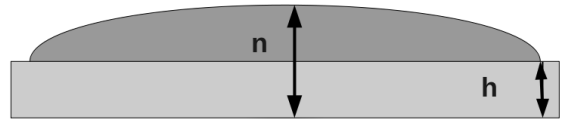}
\caption{\label{smear} Schematic diagram of the smearing profile, $h$ is the minimum number of smearing sweeps applied at the last link giving rise to smearing radius $L_{1}$, $n$ is the maximum number of smearing sweeps in the middle resulting in the radius $L_{2}$.}
\end{center}
\end{figure}
 
\noindent For a Wilson loop with $N$ spatial links, we consider a class of distributions characterized by the radii $r= f(x)+ L_1$, with an ellipsoidal  constraint
\begin{equation}
 \frac{f^{2}(x_i)}{b^{2}}+\frac{x_i^{2}}{a^2}=1
\label{elli}
\end{equation}

   The abscissa $x$ are lattice coordinates and are measured from the middle plane between the two quarks $x=0$. The shape is fixed by the minimum number of smearing sweeps at the last spatial link $n_{N}=h$ and the maximum number of sweeps at the middle links $n_{N/2}=n$.  The shape consists of a base defined by the family of rectangles of height $L_2\propto \sqrt{h}$ 
and ellipsoidal caps with $a=R/2$ and $b^2 \propto (n-h)$, thus, the radius at middle link is $L_2=b+L_1$. This parametrizes the geometrical shape schematically represented in Fig.~\ref{smear}. Among a variety of heuristic shapes, this  particular prescription is found to be especially useful for maximizing the overlap with the ground state by  variations of the tuning parameters $h$ and $n$.

\section{Numerical results and discussions}

  We take our measurements on 200 SU(3) pure-gauge configurations. The configurations are generated using the standard Wilson gauge action $S_{w}$ on a lattices of a spatial volume of $36^{3}\times 32$ for the considered  coupling value of  $\beta = 6.00$.
  
 The lattice spacing at this value is $a=0.1$ fm. The Monte Carlo updates are implemented with a pseudo-heat bath algorithm \cite{Cabibbo} using Fabricius-Haan and Kennedy-Pendelton (FHKP) \cite{Fabricius,Kennedy} updating. Each update step consists of one heatbath and 5 micro-canonical steps. The measurements are taken on configurations separated with 1000 updating sweeps.

  The APE smearing operation Eq.~\eqref{APE} and Eq.~\eqref{proj} is locally applied on spatial links of the Wilson loop with smearing parameter $\alpha=0.7$. That is, the number of smearing sweeps at each link as one moves from the quark to the antiquark is not necessarily equal. The smeared links are drawn from sets of smeared configurations corresponding to 1 to 40 sweeps of APE smearing. The spatial links in the Wilson loop are drawn from theses sets. 

  For noise reduction, the Wilson loop is calculated at each node of the lattice and then averaged over the 4-volume of the hypertoroid. The overlap with the ground state $C_{0}$ of Eq.~\eqref{over}is measured using Wilson loops of temporal extent of $2$ and $3$ slices for source separations $R=10\,a$ and $R=12\,a$.

\noindent The sequence of the numbers of smearing sweeps applied at each link labels a trial state. Here, we consider measurements of $C_{0}$ for states in the parameter  space $\{5 \leq n \leq 40,1< h <30\}$. The state is uniquely determined by $n$ and $h$. The number of sweeps at each link in between is obtained from  Eqs.~\eqref{elli} and ~\eqref{size}. The smearing profile is symmetric with respect to the middle point between the quarks.

\begin{figure}[!hpb]
\begin{center}
\includegraphics[ width=8cm]{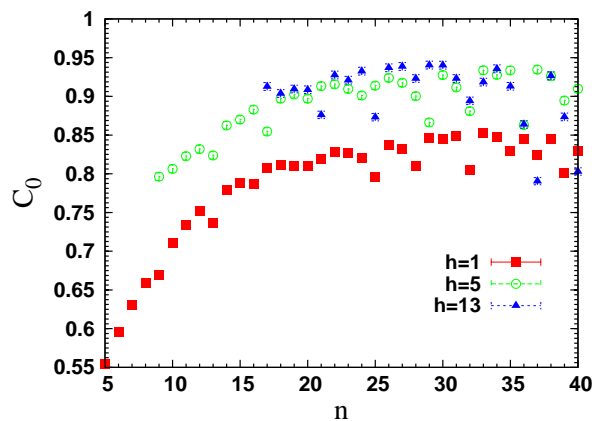}
\caption{\label{hn1}The overlap with the ground state $C_{0}$, the distance between the quark anti-quark source is $R=10\,a$, $\beta=6$.}
\end{center}
\end{figure}

   Figure~\ref{hn1} indicates the measurements of the overlap of the ground state for  three selected lines in the parametric space correspond to $\{n, h=1\}$, $\{n, h=5\}$, $\{n, h=13\}$. The shape is elliptic with the quark source at the end of the ellipse for $h=1$ (one smearing sweep at the last link). The values of $C_{0}$ are small for ellipses with number of sweeps in the middle $ 5\leq n \leq 17$. With further increase in the height $n$, the value of $C_0$ increases and lies approximately in the range $[0.80, 0.85] $ for  $n \geq 17$. However, if  the height of the rectangular shape in the base increases to $h=5$ in terms of sweeps, the overlap with the ground state is in the range $0.90 < C_{0} < 0.93$ for $n=19$ to $n=33$. Further increase of $h$ causes subtle increases in the value of $C_{0}$  until an optimum value for $h \simeq 13$ is reached. 

  The flux tube modeled as an ellipsoid with the color source lies at the far end of the prolate shaped glue does not provide the optimal overlap with the ground state in static mesons. The subsequent increase in the ground state overlap value $C_{0}$ observed by introducing this rectangular base of height $h$ can be understood as considering a model of an elliptic-like shaped flux tube in which the quark positions are shifted from the edges to the inside. The bag model, for example, leads to an ellipsoidal approximation for the gluonic field distribution around the color source  \cite{Juge1998543}. The above result would indicate that the ground state gluonic bag could have the position of quarks not exactly at the edges. Indeed the elliptical shapes revealed at small quark separations in Ref.~\cite{Bissey} contained the quarks.

%
 
\begin{figure}[!hpt]
\begin{center}
\includegraphics[ width=8cm]{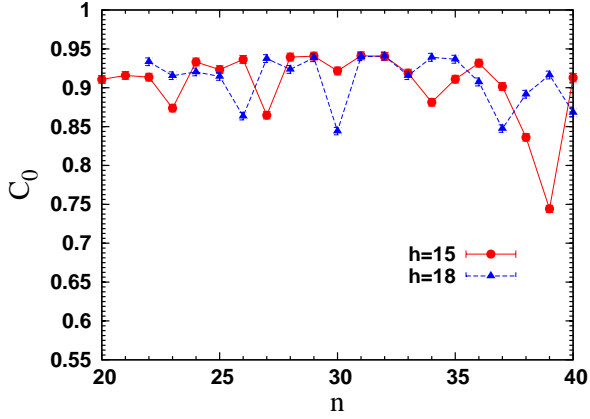}
\caption{\label{hn2}The overlap with the ground state $C_{0}$, for $R=1$ fm. The line connects the states corresponding to variation of the ellipse semi-major axis for each rectangular base corresponding to sweeps $h=15$ and $h=18$.}
\end{center}
\end{figure}

  The retrieved values of the overlap for parameter values corresponding to the two lines $\{n,h=15,18\}$ are illustrated in Fig.~\ref{hn2}. The overlap with the ground state exhibits a pronounced oscillatory  behavior versus $n$ for $ h > 13 $. Nevertheless, the measured data are not randomly scattered in the graph.

\begin{figure}[!hpt]
\begin{center}
\includegraphics[width=8cm,height=6cm]{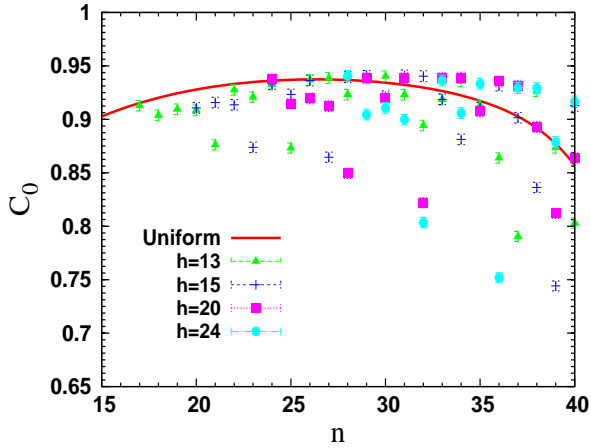} 
\caption{\label{Compare}Comparison between non-uniformly smeared profiles $n\neq h$ and uniform states $n=h$ represented by the smooth line. The smooth line connects central values of $C_{0}$ for $n=h$. The quark source separation distance $R=1$ fm.}
\end{center}
\end{figure}

  The data are seen to arrange themselves to lie ultimately within what resembles a band structure. This branching is more evident when plotting a denser region of the parametric space as in Fig.~\ref{Compare} for sweeps $n >25$. The data appear to line up into five bands with the continuous variation of $n$ indicating that this observed oscillatory behavior by changing $n$ for a given $h$ as in Fig.~\ref{hn1} may not be arbitrary. This likely to arise from the discrete nature of $n_{i}$ and $h$ in constructing the source and the sink and the inclusion of spatial link configuration which systematically probe excited states of the glue.

\begin{figure}[!hpb]
\begin{center}
\includegraphics[width=8cm,height=6cm] {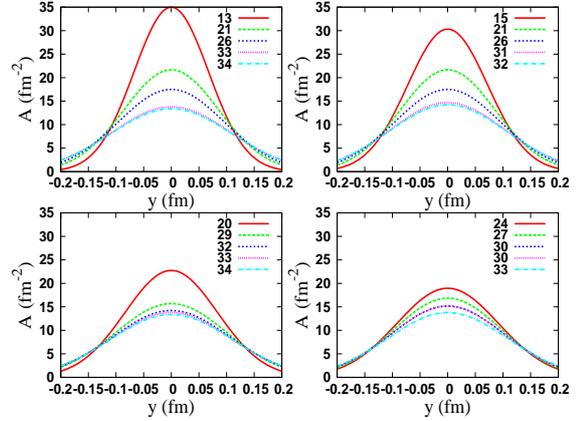} 
\caption{\label{gauss} The flux tube operator Eq.~\eqref{green}. Each operator consists of family of five Gaussian. The operators correspond to the states (h=13, n=34), (h=15, n=32), (h=20, n=34) and (h=24, n=33). Theses states maximizes the value of the with the ground state. The source separation distance $R=10$. }
\end{center}
\end{figure}

  There is, however, a variety of states of interest that maximize the overlap value at $C_{0} \approx 0.94$. These states line up in the first band from above as in Fig.~\ref{Compare}. The form of the corresponding operators for  four of these parametric states are shown in Fig.~\ref{gauss}. \\
\\
It is evident that the states with very large values of h, for example, (h=24, n=32) tend to assume a more flat shape rather than  the clear difference in amplitude along the tube as in the state (h=13, n=34). Nevertheless, the four operators overlap with the ground state equivalently.

  For comparison, the values of $C_{0}$ corresponding to the uniformly smeared (flat) states, $n=h$, is also depicted in Fig.~\ref{Compare}. In this case, $C_0$ is a smoothly varying function of $n$. In addition, the curve interestingly crosses through the states of the second band from above. Inspection of Fig.~\ref{Compare} shows that states constructed by non-uniformly smeared links can maximize the overlap with the ground state in a comparative way to the mesonic states with uniform flux tubes $\{n=h\}$. 

 In general, we do observe that neither smearing approach can overlap higher with the ground state for all the variables of the considered parametric spaces. In the far region (large values of $n$ or $h$), however, there exists many states belonging to the highest band in Fig.~\ref{Compare} for which the overlap with the ground state, $C_{0}$, is higher than the corresponding flat smearing. This observation indicates that smearing near the quark anti-quark pair, for sweeps greater than 30, increases the excited-state contamination. The links near the quark positions exhibit different UV behavior from the links at the middle with respect to the ground state overlap. The measured data provide an explanation why the usual flat APE smearing decreases the overlap with the ground state for large number of smearing sweeps.

  Recalling that the local smearing operator $\mathcal{S}^{n}U$ not only alters the width of the fat link at each locus which is proportional to $\sqrt{n}$, but also decreases the amplitude as indicated in Eq.~\eqref{green}. We see that some of these states that maximize the ground state overlap show in addition to the variation in the amplitude a sensible variation in the amplitude of the flux tube operator  near the quark source. \begin{figure}[!hpt]
\begin{center}
\includegraphics[width=8cm,height=6cm]{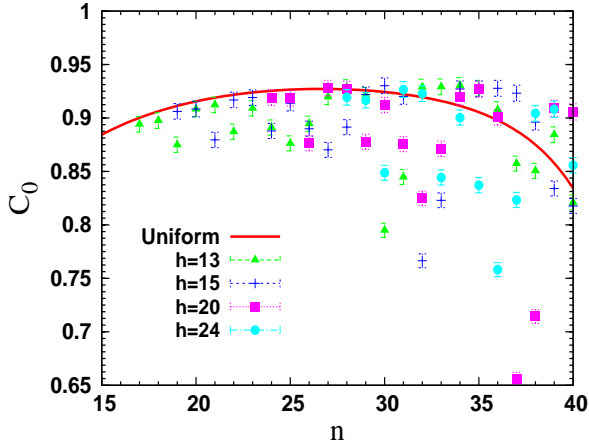} 
\caption{\label{Compare2} Same as Fig.\ref{Compare}, for quark-antiquark separation distance $R=1.2$ fm.}
\end{center}
\end{figure} \begin{figure}[!hpt]
\begin{center}
\includegraphics[width=8cm]{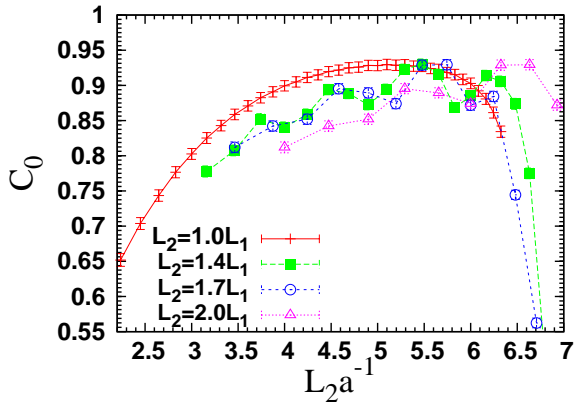}
\caption{ \label{Compare3} The overlap with the ground state $C_0$ versus the physical width of the flux-tube operator in the middle $L_{2}$ of the tube. Each line corresponds to a fixed ratio between the width of the tube in the middle and at the last link $L_1$. The quark--antiquark separation distance of $R=1.2$ fm is considered here.}
\end{center}
\end{figure}\\
\\
  At larger source separation R=1.2 fm, the collection of $C_{0}$ in branching bands is less obvious as can be seen in Fig.~\ref{Compare2}. The observable difference between the overlaps value at $R=1.2$ fm  in Fig.~\ref{hn1} and for $R=1.0$ fm in Fig.~\ref{hn2} is that the states $\{h=24,n>32\}$, having a flatter and more uniform profile, do not optimize the overlap with the ground state.

  We replot the points in the parametric space which correspond to a fixed ratio between the radii at the end and middle points of the flux tube $L_1/L_2$ in Fig.~\ref{Compare3}. The lines pass through trial states that approximately have the same non-trivial topology for the case $L_1\neq L_2 $. The flat states, $L_1=L_2$, are smoothly varying in comparison to the non-uniformly smeared states which in general assume the same behavior except for an obvious existence of fluctuation along each curve. These arise due to the fact that non-uniform states, because of the lattice structure, do not have exactly the same non-trivial topology in comparison to the flat states.     

\section{Conclusion}
   The overlap with ground state static mesons has been measured for a variety of trial mesonic states corresponding to non-uniform gluonic distributions. An optimal ground state overlap for non-uniform flux-tube operators as well as flat smeared operators has been found. This supports the possibility that the true ground state flux tube is not uniform but rather has a curved flux strength profile larger in the middle with higher action-density suppression. Such a result resembles the profile revealed at finite temperature. The findings of this work motivate the use of a Multi-level approach to explore the action-density profile of a static meson at zero temperature. This is the focus of the forthcoming investigation. It is remarkable that highly non-uniform trial states produce values of $C_{0}$ equally as good as the traditional uniform smearing approach. This result indicates that it is the smearing extent is the most critical in obtaining optimal overlap with the ground state. However, it is important to note that the shape of the non-uniform sources is critical to obtaining large values for $C_{0}$. Our investigation of other source shapes complementary to Fig.~\ref{smear} did not produce values for $C_{0}$ as large as favorable case studied here in detail.   

\section*{Acknowledgments}
  This research was undertaken on the NCI National Facility in Canberra, Australia, which is supported by the Australian Commonwealth Government.  We also thank eResearch SA for generous grants of
supercomputing time which have enabled this project. This research is
supported by the Australian Research Council.

\bibliographystyle{elsarticle-num}
\bibliography{UVref}

\begin{thebibliography}{10}
\expandafter\ifx\csname url\endcsname\relax
  \def\url#1{\texttt{#1}}\fi
\expandafter\ifx\csname urlprefix\endcsname\relax\def\urlprefix{URL }\fi
\expandafter\ifx\csname href\endcsname\relax
  \def\href#1#2{#2} \def\path#1{#1}\fi

\bibitem{luscher}
M.~Luscher, P.~Weisz, \href{http://arXiv.org:hep-lat/0207003}{Quark confinement
  and the bosonic string}, JHEP0207 049.
\newline\urlprefix\url{http://arXiv.org:hep-lat/0207003}

\bibitem{Juge:2002br}
K.~J. Juge, J.~Kuti, C.~Morningstar, {Fine structure of the QCD string
  spectrum}, Phys. Rev. Lett. 90 (2003) 161601.
\newblock \href {http://arxiv.org/abs/hep-lat/0207004}
  {\path{arXiv:hep-lat/0207004}}, \href
  {http://dx.doi.org/10.1103/PhysRevLett.90.161601}
  {\path{doi:10.1103/PhysRevLett.90.161601}}.

\bibitem{Caselle:1995fh}
M.~Caselle, F.~Gliozzi, U.~Magnea, S.~Vinti, {Width of Long Colour Flux Tubes
  in Lattice Gauge Systems}, Nucl. Phys. B460 (1996) 397--412.
\newblock \href {http://arxiv.org/abs/hep-lat/9510019}
  {\path{arXiv:hep-lat/9510019}}, \href
  {http://dx.doi.org/10.1016/0550-3213(95)00639-7}
  {\path{doi:10.1016/0550-3213(95)00639-7}}.

\bibitem{2004JHEP10005C}
M.~{Caselle}, M.~{Pepe}, A.~{Rago}, {Static quark potential and effective
  string corrections in the (2+1)-d SU(2) Yang-Mills theory}, Journal of High
  Energy Physics 10 (2004) 5--+.
\newblock \href {http://arxiv.org/abs/arXiv:hep-lat/0406008}
  {\path{arXiv:arXiv:hep-lat/0406008}}, \href
  {http://dx.doi.org/10.1088/1126-6708/2004/10/005}
  {\path{doi:10.1088/1126-6708/2004/10/005}}.

\bibitem{Bali}
G.~S. Bali, C.~Schlichter, K.~Schilling, Observing long color flux tubes in
  su(2) lattice gauge theory, Phys. Rev. D 51~(9) (1995) 5165--5198.
\newblock \href {http://dx.doi.org/10.1103/PhysRevD.51.5165}
  {\path{doi:10.1103/PhysRevD.51.5165}}.

\bibitem{Okiharu2004745}
F.~Okiharu, R.~M. Woloshyn, A study of colour field distributions in the
  baryon, Nuclear Physics B - Proceedings Supplements 129-130 (2004) 745 --
  747, lattice 2003.
\newblock \href {http://dx.doi.org/DOI: 10.1016/S0920-5632(03)02700-2}
  {\path{doi:DOI: 10.1016/S0920-5632(03)02700-2}}.

\bibitem{Bissey}
F.~Bissey, et~al., {Gluon flux-tube distribution and linear confinement in
  baryons}, Phys. Rev. D76 (2007) 114512.
\newblock \href {http://arxiv.org/abs/hep-lat/0606016}
  {\path{arXiv:hep-lat/0606016}}, \href
  {http://dx.doi.org/10.1103/PhysRevD.76.114512}
  {\path{doi:10.1103/PhysRevD.76.114512}}.

\bibitem{PhysRevD.82.094503}
A.~S. Bakry, et~al., String effects and the distribution of the glue in static
  mesons at finite temperature, Phys. Rev. D 82~(9) (2010) 094503.
\newblock \href {http://dx.doi.org/10.1103/PhysRevD.82.094503}
  {\path{doi:10.1103/PhysRevD.82.094503}}.

\bibitem{Doi2005559}
T.~Doi, N.~Ishii, M.~Oka, H.~Suganuma, The lattice qcd simulation of the
  quark-gluon mixed condensate at finite temperature and the phase transition
  of qcd, Nuclear Physics B - Proceedings Supplements 140 (2005) 559 -- 561,
  lATTICE 2004 - Proceedings of the XXIInd International Symposium on Lattice
  Field Theory.
\newblock \href {http://dx.doi.org/DOI: 10.1016/j.nuclphysbps.2004.11.341}
  {\path{doi:DOI: 10.1016/j.nuclphysbps.2004.11.341}}.

\bibitem{kac}
O.~Kaczmarek, F.~Karsch, E.~Laermann, M.~Lutgemeier, Heavy quark potentials in
  quenched qcd at high temperature, Phys. Rev. D 62~(3) (2000) 034021.
\newblock \href {http://dx.doi.org/10.1103/PhysRevD.62.034021}
  {\path{doi:10.1103/PhysRevD.62.034021}}.

\bibitem{Bakry:2010sp}
A.~S. Bakry, D.~B. Leinweber, A.~G. Williams, {Bosonic string behavior in UV
  filtered QCD}\href {http://arxiv.org/abs/1011.1380} {\path{arXiv:1011.1380}}.

\bibitem{heinzl-2008-78}
T.~Heinzl, A.~Ilderton, K.~Langfeld, M.~Lavelle, W.~Lutz, D.~McMullan,
  \href{doi:10.1103/PhysRevD.78.034504}{Is the ground state of yang-mills
  theory coulombic?}, Physical Review D 78 (2008) 034504.
\newline\urlprefix\url{doi:10.1103/PhysRevD.78.034504}

\bibitem{Allais}
A.~Allais, M.~Caselle, {On the linear increase of the flux tube thickness near
  the deconfinement transition}, JHEP 01 (2009) 073.
\newblock \href {http://arxiv.org/abs/0812.0284} {\path{arXiv:0812.0284}},
  \href {http://dx.doi.org/10.1088/1126-6708/2009/01/073}
  {\path{doi:10.1088/1126-6708/2009/01/073}}.

\bibitem{Juge1998543}
K.~J. Juge, J.~Kuti, C.~J. Morningstar, Bag picture of the excited qcd vacuum
  with static q source,, Nuclear Physics B - Proceedings Supplements 63~(1-3)
  (1998) 543 -- 545, proceedings of the XVth International Symposium on Lattice
  Field Theory.
\newblock \href {http://dx.doi.org/DOI: 10.1016/S0920-5632(97)00828-1}
  {\path{doi:DOI: 10.1016/S0920-5632(97)00828-1}}.

\bibitem{Heinzl:2007kx}
T.~Heinzl, et~al., {Probing the ground state in gauge theories}, Phys. Rev. D77
  (2008) 054501.
\newblock \href {http://arxiv.org/abs/0709.3486} {\path{arXiv:0709.3486}},
  \href {http://dx.doi.org/10.1103/PhysRevD.77.054501}
  {\path{doi:10.1103/PhysRevD.77.054501}}.

\bibitem{Albanese}
M.~Albanese, et~al., {Glueball Masses and String Tension in Lattice QCD}, Phys.
  Lett. B192 (1987) 163--169.
\newblock \href {http://dx.doi.org/10.1016/0370-2693(87)91160-9}
  {\path{doi:10.1016/0370-2693(87)91160-9}}.

\bibitem{Takahashi}
T.~T. Takahashi, H.~Suganuma, Y.~Nemoto, H.~Matsufuru, {Detailed analysis of
  the three quark potential in SU(3) lattice QCD}, Phys. Rev. D65 (2002)
  114509.
\newblock \href {http://arxiv.org/abs/hep-lat/0204011}
  {\path{arXiv:hep-lat/0204011}}, \href
  {http://dx.doi.org/10.1103/PhysRevD.65.114509}
  {\path{doi:10.1103/PhysRevD.65.114509}}.

\bibitem{Cabibbo}
N.~Cabibbo, E.~Marinari, {A New Method for Updating SU(N) Matrices in Computer
  Simulations of Gauge Theories}, Phys. Lett. B119 (1982) 387--390.
\newblock \href {http://dx.doi.org/10.1016/0370-2693(82)90696-7}
  {\path{doi:10.1016/0370-2693(82)90696-7}}.

\bibitem{Fabricius}
K.~Fabricius, O.~Haan, {HEAT BATH METHOD FOR THE TWISTED EGUCHI-KAWAI MODEL},
  Phys. Lett. B143 (1984) 459.
\newblock \href {http://dx.doi.org/10.1016/0370-2693(84)91502-8}
  {\path{doi:10.1016/0370-2693(84)91502-8}}.

\bibitem{Kennedy}
A.~D. Kennedy, B.~J. Pendleton, {Improved Heat Bath Method for Monte Carlo
  Calculations in Lattice Gauge Theories}, Phys. Lett. B156 (1985) 393--399.
\newblock \href {http://dx.doi.org/10.1016/0370-2693(85)91632-6}
  {\path{doi:10.1016/0370-2693(85)91632-6}}.

\end{thebibliography}

\end{document}